\documentclass[12pt]{article}

\usepackage[english]{babel}
\usepackage[utf8x]{inputenc}
\usepackage[T1]{fontenc}
\usepackage{wrapfig}
\usepackage{authblk}
\usepackage{cite}
\usepackage{fancyhdr}
\usepackage{subcaption}
\usepackage[compact]{titlesec}
\usepackage{enumitem}

\usepackage[a4paper,top=2.54cm,bottom=2.54cm,left=2.54cm,right=2.54cm,marginparwidth=1.5cm]{geometry}

\usepackage{amsmath}
\usepackage{graphicx}
\usepackage[colorinlistoftodos]{todonotes}
\usepackage[colorlinks=true, allcolors=blue]{hyperref}

\fancypagestyle{plain}{%
  \fancyhf{} 
  \fancyfoot[L]{\fontsize{11}{11}{\emph{This work was carried out at the Jet Propulsion Laboratory, California Institute of Technology, under a contract with the National Aeronautics and Space Administration (80NM0018D0004). \\ \copyright 2021. All rights reserved.}}}
  
}

\title{Probing the Cosmic Dark Ages with the Lunar Crater Radio Telescope}
\author{}
\date{}

\begin{document}
\maketitle
\noindent Primary author: Ashish Goel
\newline
\emph{Jet Propulsion Laboratory, California Institute of Technology}
\newline
Phone: 626-660-4714
\newline
Email: ashish.goel@jpl.nasa.gov
\newline
\newline
Co-authors: Saptarshi Bandyopadhyay\textsuperscript{1}, Joseph Lazio\textsuperscript{1}, Paul Goldsmith\textsuperscript{1}, David Bacon\textsuperscript{2}, Adam Amara\textsuperscript{2}, Steven Furnaletto\textsuperscript{3}, Patrick McGarey\textsuperscript{1}, Ramin Rafizadeh\textsuperscript{1}, Melanie Delapierre\textsuperscript{1}, Manan Arya\textsuperscript{1}, Dario Pisanti\textsuperscript{4, 5}, Gaurangi Gupta\textsuperscript{1}, Nacer Chahat\textsuperscript{1}, Adrian Stoica\textsuperscript{1}, Issa Nesnas\textsuperscript{1}, Marco Quadrelli\textsuperscript{1}, Gregg Hallinan\textsuperscript{6}, Kenneth Jenks\textsuperscript{7}, Ronald Wilson\textsuperscript{8}
\newline
\newline
\textsuperscript{1}\emph{Jet Propulsion Laboratory, California Institute of Technology}

\noindent \textsuperscript{2}\emph{Institute of Cosmology and Gravitation, University of Portsmouth, UK}

\noindent \textsuperscript{3}\emph{Physics and Astronomy, University of California Los Angeles}

\noindent \textsuperscript{4}\emph{Scuola Superiore Meridionale, Naples, Italy}

\noindent \textsuperscript{5}\emph{INFN Sez. di Napoli, Compl. Univ. di Monte S. Angelo, Naples, Italy}

\noindent \textsuperscript{6}\emph{Division of Physics, Mathematics and Astronomy, California Institute of Technology}

\noindent \textsuperscript{7}\emph{NASA Lyndon B. Johnson Space Center, Houston, TX}

\noindent \textsuperscript{8}\emph{Brigadier General, US Air Force, Retired}

\begin{figure*}[!h]
\centering
\includegraphics[width=\columnwidth]{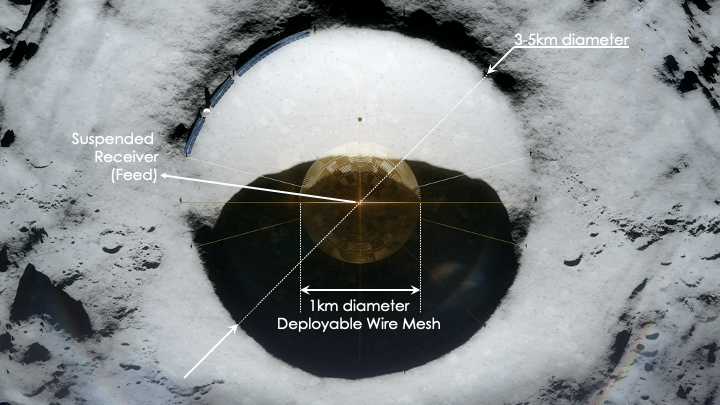}
\end{figure*}

\pagebreak

\begin{abstract}
The Cosmic Dark Ages represent the period in the early evolution of the Universe, starting immediately after the decoupling of CMB photons from matter, and ending with the formation of the first stars and galaxies. The H{\scriptsize I} signal from the neutral hydrogen atoms is the only mechanism for us to understand this crucial phase in the cosmological history of the Universe and answer fundamental questions about the validity of the standard cosmological model, dark matter physics, and inflation. Due to cosmological redshift, this signal is now only observable in the 3-30~MHz frequency band, which is blocked from reaching the surface of the Earth by the ionosphere. In this paper, we present the design of the Lunar Crater Radio Telescope that intends to carry out unprecedented measurements of this signal by deploying a kilometer-sized parabolic reflector mesh inside a lunar crater on the far side of the Moon and suspending a receiver at its focus.  
\end{abstract}

\section{Introduction}
\label{sec.Intro}

Shortly after the Big Bang, the Universe expanded and cooled enough to allow the protons and electrons to combine, forming neutral hydrogen. The associated decoupling of photons and matter allowed the photons to travel freely through the Universe, constituting what is observed today as the cosmic microwave background (CMB) radiation. For millions of years after this `recombination' event, the Universe was devoid of any sources of illumination. This period is known as the `Cosmic Dark Ages'. As the Universe continued to expand, density fluctuations grew and gravitational collapse led to formation of the first luminous sources, during a period known as the Cosmic Dawn. 


Understanding how our Universe transitioned from a gas comprising mostly neutral hydrogen, to a collection of stars and galaxies, is one of the most fundamental questions in cosmology. But due to the dearth of electromagnetic sources, we have been devoid of observations that can shed light into the evolution of the Universe during this period. The only source of information is the H{\scriptsize I} ($\lambda = 21$ cm, $\nu = 1420$ MHz) signal associated with the spin-flip transition in neutral Hydrogen. Due to cosmological redshift, this 21~cm signal from the Dark Ages is currently visible at ultra-long radio wavelengths of 10~m or higher. Observations of these signals cannot be carried out from the surface of the Earth due to absorption from the Earth's ionosphere. In this paper, we present the design of a telescope that can be robotically assembled by suspending a wire mesh inside a lunar crater on the far side of the Moon, to carry out unprecedented observations of the signal from the Cosmic Dark Ages. This telescope is referred to as the Lunar Crater Radio Telescope (LCRT). Being on the far side of the Moon offers the added benefit of isolating the telescope from sources of radio interference on Earth. 

\section{Science Objectives}
\subsection{Fundamental Science Questions Addressed by LCRT}
\label{sec.Science}

The scientific benefits of observing the 21~cm signal from the Dark Ages have been captured thoroughly by Burns \emph{et al.}\cite{burns2019dark} and Furlanetto \emph{et al.}\cite{Ref:Furlanetto2019_Astro_White_paper}. While Burns \emph{et al.} describe the fundamental science questions that can be answered by studying the sky-averaged absorption spectrum of the 21~cm signal, Furnaletto \emph{et al.} describe what can be learned by studying the spatial fluctuations in this signal. We briefly reproduce these scientific highlights and the reader is encouraged to peruse the white papers, and associated references, for a more thorough description. 

\begin{wrapfigure}{R}{0.5\textwidth}
	\centering
	\includegraphics[width=0.45\textwidth]{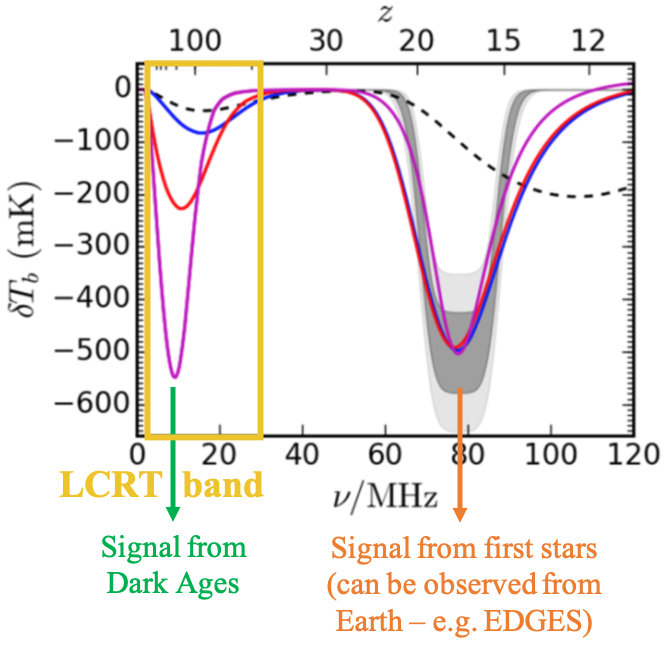}
	\caption{Signals from Dark Ages and First Stars ``Cosmic Dawn'' (image credit:~\cite{burns2019dark}) }
	\label{fig:Furlaneto_signal}
\end{wrapfigure}

During the Dark Ages, the Universe was considerably simpler, consisting mainly of neutral hydrogen, photons, and dark matter. It therefore serves as an excellent laboratory for testing our fundamental theories of cosmology, dark matter physics and inflation. Fig.~\ref{fig:Furlaneto_signal} shows our best understanding of this signal, as a function of  cosmological redshift ($z$) or frequency ~\cite{burns2019dark}. Going towards the left on the x-axis in this plot is equivalent to going further back in time. Signals from the Dark Ages are expected to correspond to frequencies $<20$~MHz and those from the First Stars are in the 60--100~MHz band. The dotted line is based on the $\Lambda CDM$ cosmological model. The y-axis represents the relative brightness of the radio sky with respect to the CMB. 

Recent measurements using the EDGES instrument have suggested potential constraints on  the signals from the first stars~\cite{Ref:Monsalve17,Ref:Bowman2018_EDGES}. This region is highlighted in gray. Different models (shown in different colors) have been proposed to explain the extra cooling that would justify the deeper trough observed by EDGES in the $15 < z < 20$ region. A few models explain the depth of the trough by invoking a stronger background sourced by synchrotron emission from the early star-forming regions, or dark matter annihilation. A second explanation is provided by a possible change in cosmological parameters. Alternatively, the trough could also be explained by the presence of a new kind of warm dark matter that cools the neutral hydrogen through Rutherford scattering. If true, this would imply the existence of `charged' dark matter that can interact electromagnetically with baryonic matter. This striking hypothesis has broad implications on our fundamental understanding of dark matter physics. While these three models explain the deeper trough observed during Cosmic Dawn, their predictions vary widely for the Dark Ages absorption trough. Therefore, direct measurements of the sky-averaged absorption spectrum from the Dark Ages can uniquely determine which model fully explains the cooling behavior of the Universe, all the way from the beginning of Dark Ages to the Epoch of Reionization. Any observed deviations from the standard $\Lambda CDM$ cosmological model would necessitate new physics to explain baryonic cooling beyond simple adiabatic expansion of the Universe. 

In addition to the key questions answered by studying the sky-averaged 21~cm signal, several fundamental insights can be gained from the power spectrum of the spatial fluctuations in the signal. Our current understanding of the cosmological history has been driven by studying the fluctuations in the CMB signal. However, the CMB signal only allows us to probe a narrow region around recombination. On the other hand, the spectral shape of the 21~cm signal allows us to probe all three spatial dimensions and study how these fluctuations evolved over time. Further, it allows us to expand the dynamic range of power spectrum measurements since fluctuations during the Dark Ages vary over a wider range of length scales. Specifically, studying the spatial structure of the 21~cm signal will support the following investigations
\begin{itemize}[leftmargin=*]
\itemsep0em 
	\item Since the power spectrum depends on various cosmological parameters, measurements of fluctuations in the 21~cm signal allows us to establish stringent constraints on the values of key parameters such as the total spatial curvature, neutrino masses and the running of the spectral index of the matter power spectrum \cite{Ref:MaoTegmark2008}.
	\item A key parameter in understanding the origin of density fluctuations in the Universe is the level of non-Gaussianity at the end of inflation. Measurements from CMB do place limits on this parameter but are contaminated by the non-Gaussianity introduced by structure formation in the Universe. 21~cm measurements serve as a clean mechanism to probe various models of cosmological inflation and test the inflationary hypothesis in general \cite{Ref:Chen2018Inflation}.
\end{itemize}

\subsection{Secondary Science Objectives}
\textbf{Radio Emissions from Extrasolar Planets}: Generated by dynamo processes within the planet, planetary-scale magnetic fields are a remote-sensing method to constrain the properties of a planet's interior. In the case of the Earth, its magnetic field has also been speculated to be partially responsible for its habitability. Thus, knowledge of the magnetic field of an extrasolar planet may be a necessary component of assessing its habitability, or understanding an absence of life on an otherwise potentially habitable planet.  The LCRT's beam will pass across some of the known extrasolar planets, as shown in Fig.~\ref{Fig:CombinedCraterSelection}(a), and LCRT might be able to observe these radio emissions. 

\textbf{Extension of LCRT's Wavelength Range}: It might be possible to increase LCRT's wavelength (frequency) range to include the 1--5~m wavelength band (i.e., 30--150~MHz frequency band) with an upgrade to the reflector's design and an additional set of receivers for these frequencies. 
The high-density mesh could be added only to the sections closer to the center, thereby not incurring a massive weight penalty. This would open the door to calibrating LCRT using observations of the same astronomical sources that are visible from Earth and further refine the ionospheric models that are the main limiting factor in conducting observations from Earth in these frequency bands. For example, the EDGES data~\cite{Ref:Monsalve17,Ref:Bowman2018_EDGES} shown in Fig.~\ref{fig:Furlaneto_signal} has come under considerable scrutiny due to their ionospheric corrections. 

Finally, this would also enable very large baseline interferometry (VLBI) between LCRT and Earth-based radio telescopes.
This might open the door to high-resolution imagery of first stars and early galaxies in these ultra-long wavelength bands.

\section{Proposed Instrument}
The LCRT project aims to build a kilometer-sized telescope by deploying a wire mesh inside a 3-5~km lunar crater on the far side of the Moon to serve as the reflector, and suspending a receiver above the mesh at the focus. 
A detailed description of the telescope design can be found in \cite{Ref:LCRT_IEEE_2021}.

\subsection{Lunar Crater Selection}

\begin{figure}
	\centering
	\includegraphics[width=0.8\textwidth]{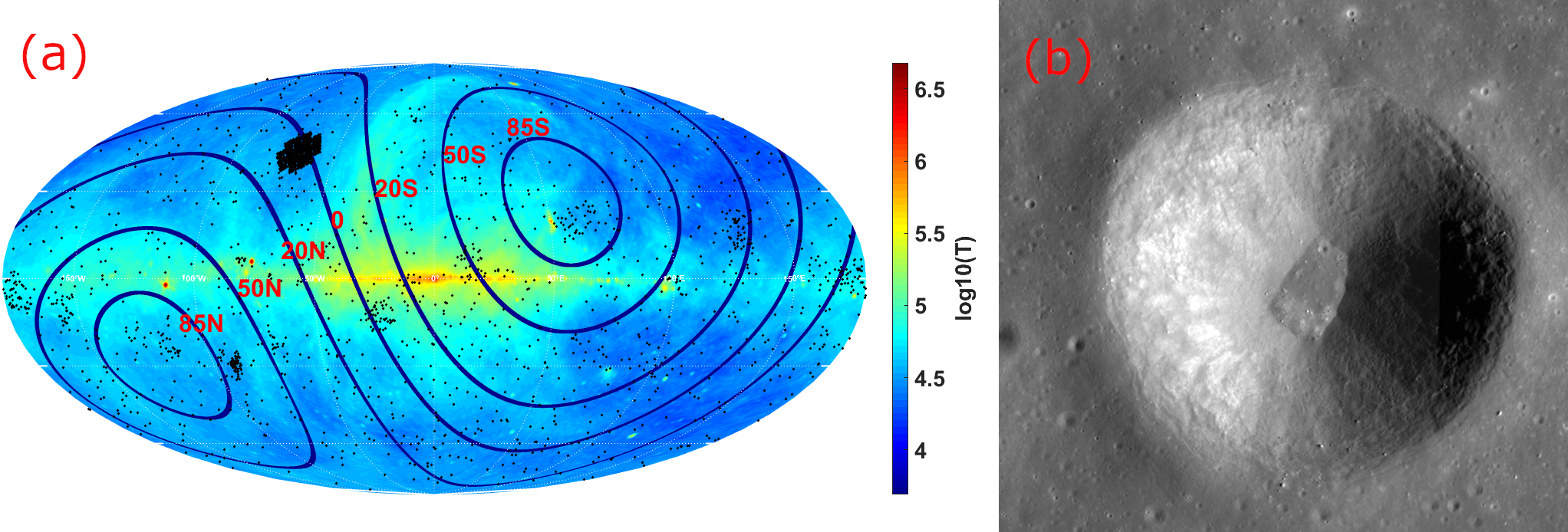}
     \caption{(a) Celestial track of the LCRT beam as a function of Lunar latitude, with known exoplanets (b) High-resolution image of the chosen crater for LCRT deployment.}
    \label{Fig:CombinedCraterSelection}
\end{figure}

In order to choose the ideal location for the telescope on the far side of the Moon, we computed the celestial track of the telescope beam as a function of its latitude on the Moon. The result of this analysis is depicted in Fig. \ref{Fig:CombinedCraterSelection}(a). Placing the telescope close to $20^{\circ}$N latitude achieves a good balance of not being overwhelmed by synchrotron radiation from the galactic center and covering a wide swath of the celestial sky. 

We manually surveyed $\approx$300 craters in the vicinity of $20^{\circ}$N latitude and $180^{\circ}$E longitude that met the following criteria
1) Diameter range within 3-5~km
2) Crater depth $>$ 600~m, so that both reflector and feed are suspended below the rim 
3) No boulders or outcrops, that could cause difficulty while deploying the reflector
4) Complete crater rim for uniform deployment of wires and anchors in all directions
The best crater chosen based on these criteria is shown in Fig. \ref{Fig:CombinedCraterSelection}b.

\subsection{Reflector Design}
Designing the reflector of the telescope is one of the primary challenges for LCRT. The reflector has to meet electromagnetic, structural and thermal requirements, while minimizing launch mass and volume. 
The current design philosophy is to make the mesh out of thin aluminum wire, and use non-conductive structural elements to provide strength and support to the structure. In order to verify this approach, preliminary numerical analysis has been carried out to determine the shape of the proposed reflector under lunar gravity. Origami-inspired techniques, similar to Starshade~\cite{Ref:Arya2020_Starshade}, for packaging and deploying the reflector, have also been analyzed using finite element methods. 
From a thermal perspective, large temperature swings of up to 300~K are observed on the surface of the Moon but the temperature stays stable within $~$10~K during the lunar night \cite{Ref:Williams2017global}. The effect of thermal expansion on the telescope performance can be minimized by restricting science operations to the lunar night when we are also shielded from solar RF noise. 

The mesh will comprise wires spaced by distances no greater than $\lambda$/4 in order to ensure high reflectivity at our smallest wavelength $\lambda$ of 10~m. A log periodic antenna is currently being designed to serve as the feed for the telescope due to its large bandwidth and stable gain pattern. 

\subsection{Concept of Robotic Operations}
\begin{figure}[!h]
	\centering
	\includegraphics[width=0.75\columnwidth]{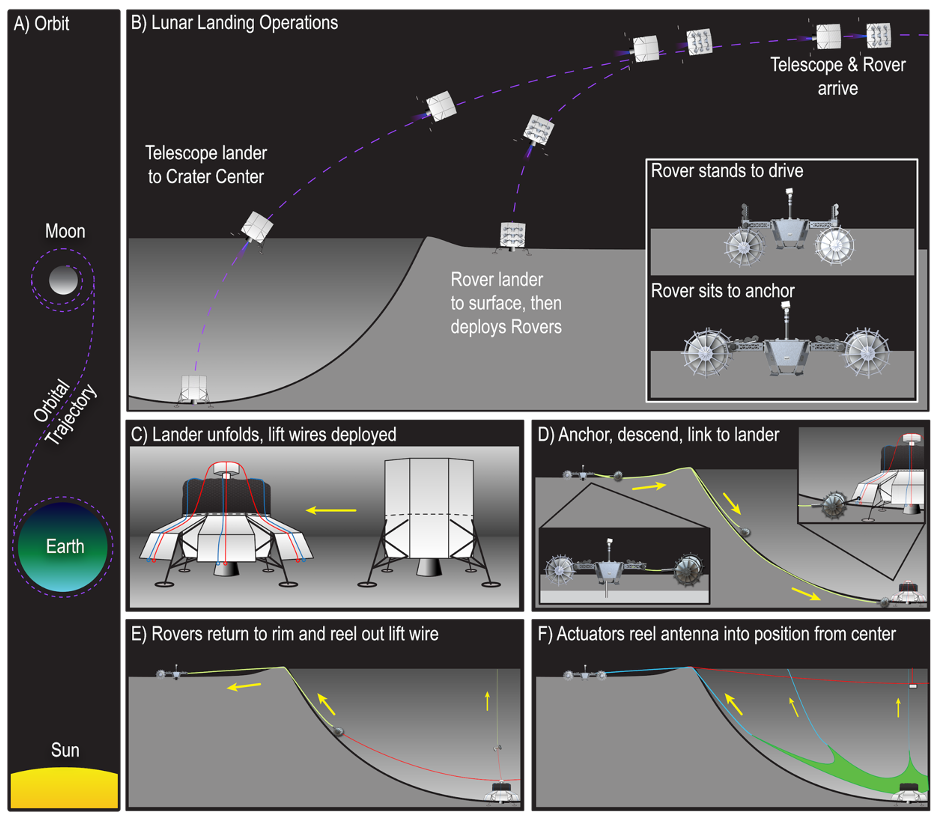}
	\caption{Concept of Operations for building LCRT.}
	\label{fig:Conops-image}
\end{figure}

The concept of operations (ConOps) for constructing LCRT is shown in Fig.~\ref{fig:Conops-image}.
One half of the spacecraft with different components of LCRT lands on the crater floor, carrying the wire mesh and the receiver. The other half lands on the crater rim, carrying DuAxel robots \cite{Ref:Mcgarey2019_DuAxel} \cite{Ref:Nesnas2019_Moon_Diver} and supporting equipment. 
The lander on the crater floor anchors itself and deploys guide wires. 
One half of each DuAxel robot anchors itself on the rim while the other half travels down the crater wall, links to one of the guide wires, and carries it back to the rim of the crater for anchoring. Once all the guide wires have been anchored, the receiver antenna is hoisted up, followed by the deployment of the reflector mesh. Further details of the deployment process are presented in \cite{Ref:McGarey2020_LCRT_Deployment}.

\section{Summary and Recommendations}
We have highlighted the revolutionary impact that H{\scriptsize I} line observations, at redshifts corresponding to the Cosmic Dark Ages, can have on our understanding of some of the most fundamental aspects of physics. We have also presented an overview of LCRT for observing this signal. There are other complimentary \emph{sparse dipole array concepts} such as the Lunar FARSIDE ~\cite{Ref:Burns2019farside}; \emph{lunar-orbiting satellite mission concepts} like 
LORAE ~\cite{Ref:Burns90lunar}, 
DARE ~\cite{Ref:Burns11_DARE}; and
\emph{multi-satellite mission concepts at the Earth-Moon L2 Lagrange point} like 
ALFA ~\cite{Ref:Jones98astronomical}, 
FIRST \cite{Ref:Bergman09first} and
OLFAR ~\cite{Ref:Rajan16space}. 
There is a great push from NASA for greater utilization of the lunar surface and for fundamental science investigations. This presents an opportune moment for using the technological infrastructure and launch opportunities for furthering our understanding of the cosmological history of the Universe. In this vein, we present the following recommendations 
\begin{enumerate}[leftmargin=*]
\itemsep0em
	\item Dark Ages cosmological studies from the lunar surface should be given high priority.
	\item Research should be funded for raising the TRL of reflector and receiver designs that can operate in the 10-100~m wavelength range under lunar environmental conditions.
	\item Research should be funded for understanding the geotechnical, dielectric, electrostatic and dynamic properties of the lunar regolith. This knowledge is critical for developing anchoring mechanisms and predicting the performance of RF instruments on the Moon.
	\item NASA should fund appropriate Earth analogue studies to demonstrate the feasibility and efficacy of deploying such instruments on the lunar surface.
	\item The far side of the Moon should be declared a radio-quiet zone so that opportunities for conducting radio science investigations from the lunar surface are not hampered by the deployment of communication and navigation infrastructure in lunar orbits.
\end{enumerate}  

\bibliographystyle{alpha}
\bibliography{BPSWhitePaper}

\end{document}